\documentclass{article}
\usepackage{spconf,amsmath,graphicx}

\usepackage{xcolor}
\usepackage{multirow}
\usepackage{booktabs}

\title{Towards Robust Speaker Verification with Target Speaker Enhancement}
\name{Chunlei Zhang, Meng Yu, Chao Weng, Dong Yu}
\address{Tencent AI Lab, Bellevue, WA, USA \\
\small \tt \{cleizhang, raymondmyu, cweng, dyu\}@tencent.com}
\begin{document}
\ninept
\maketitle
\begin{abstract}
This paper proposes the target speaker enhancement based speaker verification network (TASE-SVNet), an all neural model that couples target speaker enhancement and speaker embedding extraction for robust speaker verification (SV). Specifically, an enrollment speaker conditioned speech enhancement module is employed as the front-end for extracting target speaker from its mixture with interfering speakers and environmental noises. Compared with the conventional target speaker enhancement models, nontarget speaker/interference suppression should draw additional attention for SV. Therefore, an effective nontarget speaker sampling strategy is explored. 
To improve speaker embedding extraction with a light-weighted model, a teacher-student (T/S) training is proposed to distill speaker discriminative information from large models to small models. Iterative inference is investigated to address the noisy speaker enrollment problem. We evaluate the proposed method on two SV tasks, i.e., one heavily overlapped speech and the other one with comprehensive noise types in vehicle environments. Experiments show significant and consistent improvements in Equal Error Rate (EER) over the state-of-the-art baselines. %For better nontaget speaker suppression in SV, which is an additional constraint compared with the conventional target speaker enhancement models, a simple and effective training strategy is explored in the target speaker enhancement module.%                              
\end{abstract}

\section{Introduction}

Deep neural network (DNN) based speaker embedding models outperform the conventional models (e.g., i-vector) in many speaker verification (SV) tasks \cite{dehak2010front,snyder2018x,zhang2017end,li2017deep,zhang2018text,liu2019large,villalba2019state}, becoming the state-of-the-art paradigms of speaker recognition technology. The improved performance can be attributed to three major reasons: a) deep (sometimes very deep) neural network architectures that can effectively model speaker discriminative information from utterances \cite{snyder2018x,zhang2018text,villalba2019state,chung2019voxsrc}; b) novel neural network training objectives (e.g., large margin softmax losses, contrastive learning based losses or end-to-end losses etc.) and the corresponding training strategies \cite{liu2019large,chung2019voxsrc,zhang2018text,heigold2016end,wan2018generalized}; c) sufficient training data distributed across diverse acoustic conditions \cite{snyder2018x}.

Despite the success achieved, current SV systems still suffer evident performance degradation in the presence of severe background noise with interfering speakers \cite{zhang2019utd,Sadjadi2019,Lee2019,Shon2019,Rao2019}. Therefore, the SV system is at increased risk of failure when exposed to unconstrained conditions. Front-end techniques to tackle against such adverse environments include speaker diarization, noise suppression, and speech separation. In \cite{snyder2019speaker}, it is suggested that speaker diarization works well for the multi-speaker scenario where overlapped speech is not the main focus. However, such system fails when highly overlapped speech exists. In addition, the system performance heavily relies on speaker diarization accuracy, which remains to be challenging at high noise levels. On the contrary, a noise suppression model might perform well for general noise reduction, however, the speaker of interest from multi-speaker utterances could not be easily extracted through such models. The speech separation technologies also provide a potential solution for extracting target speaker from the mixture. The major drawback lies in the model inference, where the number of speakers has to be known in advance. In practice, such information is usually unknown from a test utterance. Due to those constraints, successful adoption of the aforementioned front-end techniques are rarely seen in generic SV systems. 

In this study, we propose the target speaker enhancement based speaker verification network (TASE-SVNet), a SV framework that is robust against background noise and interfering speakers. For speech recognition models with target speaker enhancement \cite{Wang2019,Wang18d,Delcroix18}, speaker profile utterances require additional cost to collect. While in SV tasks, enrollment utterance of a speaker is naturally given, which could be utilized as the anchor/profile signal for target speaker enhancement. Specifically, following our previously formulated time-domain target speaker enhancement model \cite{ji2020}, the joint training strategy of target speaker embedding extraction and speech enhancement is expected to offer well-performed target speaker extraction for ``target" test utterance \footnote{``Target" test utterance refers to the test utterance of a target trial, where speaker of interest is included in the test utterance. ``Nontarget" test utterance refers to the test utterance of a nontarget trial, where the speakers in the test utterance are not the enrolled speaker.}. Furthermore, to encourage the speech enhancement module to suppress noise and nontarget speakers for "nontarget" test utterance, a novel nontarget sample training strategy is introduced as an additional constraint to the target speaker enhancement training, which effectively prevents the increase of False Alarm Rate (FAR) in the SV phase. Regarding the SV model in the TASE-SVNet framework, a teacher-student (T/S) training scheme is investigated to improve the overall SV performance with a light-weighted model. To put it together, the target speaker enhancement module and the SV model of TASE-SVnet are unified in a multi-stage training procedure. 

Integrating target speaker enhancement to robust SV is still at its early stage. The most relevant work to us is \cite{Rao2019}, where speaker-beam based target speaker enhancement is evaluated \cite{Delcroix18}, and an i-vector model is employed as the SV model. In contrast with the pipeline system \cite{Rao2019}, we expect an all-neural network solution with joint training leads to improved SV performance. 

The paper is structured as follows. Sec.\ref{TASESVNet} describes the overall TASE-SVnet framework with a time-domain target speaker enhancement module, a speaker embedding module, and the improved training for each module.  Sec.\ref{experiments} details the experimental setup and the corpora. Sec.\ref{results} presents experimental results and analysis. Finally, we conclude the paper in Sec.\ref{conclude}.

\section{Speaker verification with target speaker enhancement}
\label{TASESVNet}
In this section, we firstly introduce the overall framework of TASE-SVNet. Then, we provide our improved training strategies of the target speaker enhancement module and the speaker embedding module in the following subsections. Lastly, we summarise the multi-stage training procedure of TASE-SVNet.   
\begin{figure}[t]
  \centering
  \includegraphics[width=85mm, height=42mm]{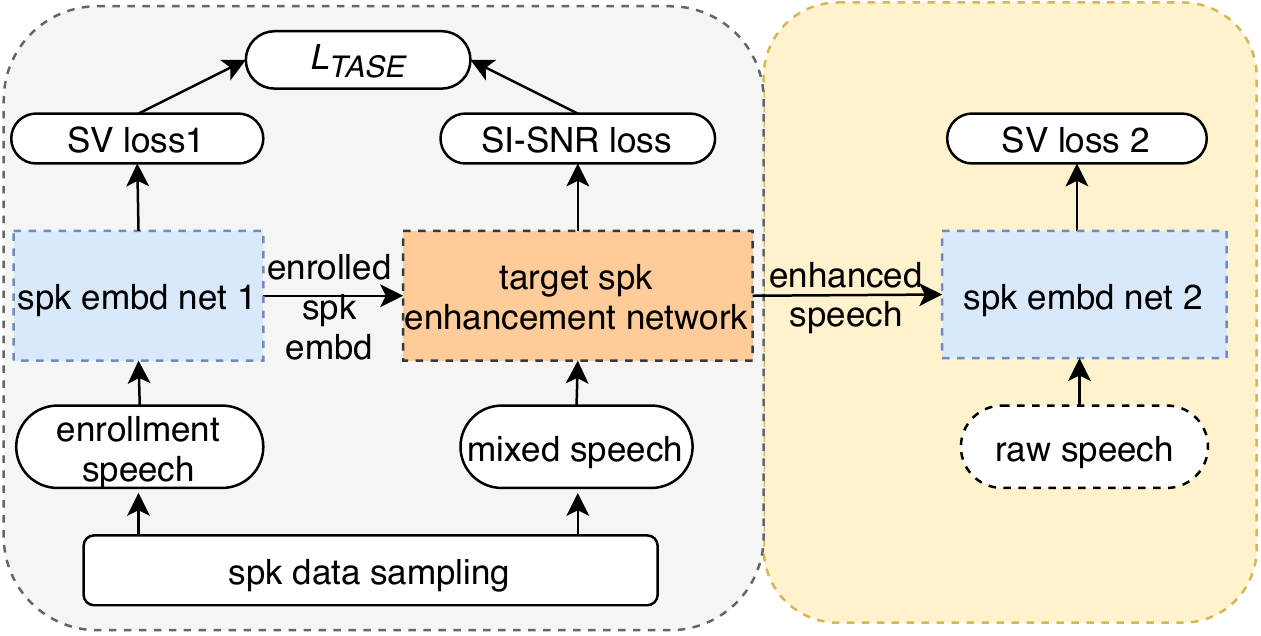}
  \vspace{-0.4cm}
  \caption{The flow diagram of TASE-SVNet framework.}
  \label{baseline}
 \vspace{-0.6cm}
\end{figure}
\subsection{The TASE-SVNet framework}
\label{system_architect}
The overall TASE-SVNet framework consists of a target speaker enhancement module and a speaker verification module. Fig.\ref{baseline} illustrates the base framework of the proposed TASE-SVNet. The target speaker enhancement module in the left part of Fig.\ref{baseline} is inherited from one of our recently developed models, which gives the best speech enhancement performance in \cite{ji2020}. Scale-invariant signal-to-noise ratio (SI-SNR) \cite{ji2020,luo2018tasnet} is used as the target speaker enhancement objective function, which is defined as:
\begin{equation} 
    \text{SI-SNR} = 20 \log_{10} \frac{\Vert \alpha \cdot \mathbf{s}_e \Vert}{\Vert \mathbf{s}_t - \alpha \cdot \mathbf{s}_e \Vert}, \label{eq1}
\end{equation}
where $\mathbf{s}_e, \mathbf{s}_t$ are estimated signal and target signal, respectively, for which zero-mean normalization is applied. $\alpha$ is an optimal scaling factor computed via $\alpha = \mathbf{s}_e^T\mathbf{s}_t / \mathbf{s}_t^T\mathbf{s}_t$. 

\begin{figure}[t]
  \centering
  \includegraphics[width=80mm, height=53mm]{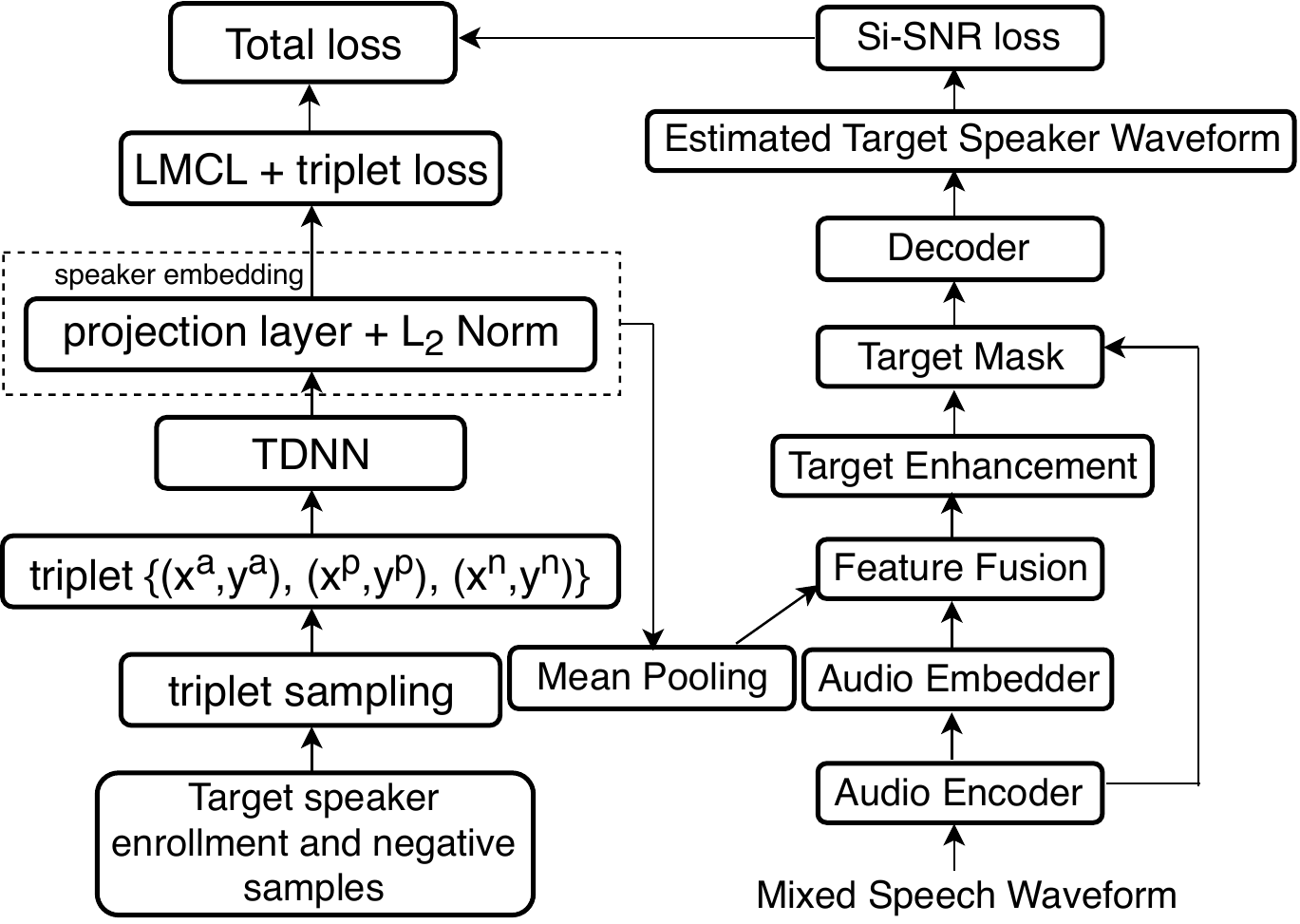}
  \vspace{-0.4cm}
  \caption{Joint learning for target speaker enhancement.}
  \label{joint}
  \vspace{-0.5cm}
\end{figure}

Fig.\ref{joint} depicts the detailed implementation of joint learning architecture, where a modified time-domain Conv-TasNet structure is employed as the speaker enhancement network and a TDNN architecture (i.e., spk embd net 1) is utilized to provide the bias information from speaker of interest (i.e., meaning pooling over enrollment speaker embeddings), respectively. To jointly learn speaker embedding and target speaker enhancement, a multi-task loss $\mathcal{L}_{TASE}$ is formulated with equal weight combination of the target speaker enhancement loss $\mathcal{L}_{SI-SNR}$ and the speaker verification loss $\mathcal{L}_{SV}$. Furthermore, $\mathcal{L}_{SV}$ is also a multi-task objective which consists of a large margin cosine loss $\mathcal{L}_{lmc}$ (LMCL) \cite{wang2018cosface} 
and a triplet loss $\mathcal{L}_{triplet}$ \cite{zhang2017end} with a tuned weight. The total loss is defined as follows:
 \begin{equation}
            \mathcal{L}_{SV} = \mathcal{L}_{triplet} + \omega_1 \mathcal{L}_{lmc} + \omega_2\mathcal{L}_{r}, \label{eq3}
\end{equation} 
where $L_r$ is a $L_2$-regularization term which alleviates over-fitting during training, $\omega_1$ and $\omega_2$ are the weights determined in the experiments. We refer the readers to \cite{ji2020} for more detailed network configuration and training procedures.   

For the SV model in the right part of Fig.\ref{baseline} (i.e., spk embd net 2), two network architectures are introduced as the benchmarks for evaluating SV performance. Specifically, a TDNN model (the same as spk embd net 1) and a more heavy-weighted Resnet-50 are investigated \cite{Chung2018,he2016deep} as contrastive systems. Both speaker embedding networks (i.e., spk embd net 1 \& 2) in Fig.\ref{baseline} are pre-trained on a fairly large data set particularly for the task of SV. More details about the experimental setups can be found in Sec. \ref{experiments}. 

\subsection{Nontarget sample training in TASE-SVNet }
\label{negative_samp}
To the best of our knowledge, all enrollment speaker informed target speaker enhancement models~\cite{Wang18d,Delcroix18,Wang2019,ji2020} are only evaluated on one condition, which assumes target speaker is $always$ presented in the test utterance. This is a strong prior that does not hold for SV tasks. We argue that the current target speaker enhancement training criterion is inclined to make the network output of nontarget test utterance more biased to enrolled speaker, leading to an increased FAR in terms of SV performance. Practically, we find our proposed base target speaker enhancement models occasionally fail to suppress interference speakers in the nontarget test utterance. 

To address this issue, we propose to add nontarget samples to the target speaker enhancement training. Specifically, a triplet consisting of enrollment speaker utterances, a nontarget test utterance and a NULL reference signal, is defined as the nontarget sample in our study. With the negative samples randomly added in model training, the target speaker embedding module is expected to have better performance in interference speaker suppression. To unify the nontarget sample training with the conventional ones (only with target samples),  we update the SI-SNR loss in Eq.(\ref{eq1}) as a new training objective for the wohle TASE-SVnet:
\begin{equation} \label{sisnr_undate}
    \text{SI-SNR} = 20 \log_{10} \frac{\Vert \alpha \cdot \mathbf{s}_e \Vert}{\Vert \mathbf{s}_t - \alpha \cdot \mathbf{s}_e \Vert}, \left\{
\begin{aligned}
\mathbf {s}_t  & = \mathbf{x}_t   & tar \\
\mathbf {s}_t & \sim N(0,\sigma^2)  & non  
\end{aligned}
\right.
\end{equation}where $\mathbf{x}_t$ is the zero-mean normalized target reference signal, and the NULL reference signal in nontarget triple is set to follow a zero-mean normal distribution, with $\sigma$=1e-6 as the stand deviation. Ideally, the NULL reference signal should be an all zero sequence. However, $\mathbf {s}_t  = \mathbf{0} $ will lead to zero loss, the gradient from nontarget sample will not apply to model update. Therefore, our solution is to provide a small-valued Gaussian disturbance.

\subsection{T/S training for speaker embedding training }
\label{TS_training}
T/S learning has been proved to be effective for domain adaptation in speech processing \cite{li2018developing,jung2019short}. By T/S training, it is expected to learn a student model that can perform well in target domain with domain specific knowledge distillation from a well-trained source-domain teacher model. 

In this study, we aim to further improve the small TDNN model with the large Resnet-50 model with T/S training. The T/S training objective is fairly simple. The pre-trained Resnet-50 model will act as the teacher model, the output of student TDNN model (i.e., speaker embedding) is forced to be close to that from the teacher model. To achieve this, we propose to use mean squared error (MSE) as the T/S training objective:
\begin{equation}
   \mathcal{L}_{MSE} = \frac{1}{N}\frac{1}{D}\sum_{i=1}^{N}\sum_{j=1}^{D}(Y_{t}(i,j)-Y_{s}(i,j))^2,
\end{equation}
where $D$ is the dimension of speaker embedding, $N$ is the number of speaker embeddings in one training batch, $Y_t$ and $Y_s$ represent the speaker embeddings from the teacher model and the student model, respectively. Theoretically, T/S learning can be performed in an unsupervised manner. In our experiments, we find that convergence of the student TDNN model can be achieved much faster with a joint SV loss described in Eq.\ref{eq3}, which requires labeled training data. The overall training loss $\mathcal{L}_S $ for the student model is as follows:
\begin{equation}
    \mathcal{L}_{S} =  \mathcal{L}_{SV} + \mathcal{L}_{MSE},
\end{equation}

\subsection{A multi-stage training procedure in TASE-SVNet}
\label{multi-stage}
Although all the modules of the TASE-SVNet framework can be jointly optimized from scratch. However, not much benefits have been seen with the end to end training. In this section, a road map to develop the final system is presented, where different modules are unified with a multi-stage training strategy, resulting in consistent system performances with less data constraints. More specifically, the multi-stage training procedure is summarized as:
\begin{itemize}
    \item[1)] Pre-train speaker embedding models (``spk embd net" 1 \& 2) with a large scale general domain speaker recognition corpus. Apply T/S training to the TDNN model, employ the T/S learned TDNN model as the initialization of ``spk embd net".
    \item[2)] Jointly train spk embd net 1 (i.e., initialized with T/S trained TDNN) and target speaker enhancement network (from scratch) with a target speaker enhancement corpus.
    \item[3)] Fix the target speaker enhancement module when training has converged. Fine tune the spk embd net 2 using the triplet loss with the enhanced and raw speech data. 
\end{itemize}

\section{Corpora and experimental setup}
\label{experiments}
\subsection{Speech corpora}
\subsubsection{SV corpora}
Two datasets are utilized to examine our base SV models. The first one is a Mandarin language corpus. The training set contains of 8800 speaker drawn from two gender-balanced public speech recognition datasets\footnote{http://en.speechocean.com/datacenter/details/254.htm}. The training data is then augmented 2 folds to incorporate variabilities from distance (reverberation),  channel or background noise, resulting in a training pool with 2.8M utterances. The duration ranges from 0.5s to 13.2s, with the average duration of 3.8s. Two real SV test sets are examined. Test set 1 is a 44-speaker corpus specifically for multi-speaker SV evaluation (noted as ``multi-spk"), each utterance may include 1-3 speakers with or without overlapping. We generate 10K trials from the ``multi-spk" test set, among which 1K trails are target trials. For speaker enrollment, 3 utterances are randomly selected, where only single-speaker utterance is used for enrollment. Test set 2 is a 365-speaker dataset collected under the unconstrained vehicle environments (noted as ``vehicle-spk"). It is a much larger dataset that covers most of the noise types (i.e., wind noise, mechanical noise, background music/noise, and interference speakers etc.). Similar as the ``multi-spk" for trial generation, we randomly select 3 utterances for speaker enrollment, and sample the rest for test. 100K trials are produced in the end, among which 10K trials are target trials. For system comparisons, we also report our SV baselines on the Voxceleb task \cite{Chung2018,chung2019voxsrc}. The performance is reported on Voxceleb1 test trials.

\subsubsection{Target speaker enhancement corpus}
For training the target speaker enhancement model, we use the same corpus as \cite{ji2020}, which is simulated from AISHELL-2 \cite{Du2018}. The whole dataset is split to a 1900-speaker training set, a 60-speaker validation set and a 30-speaker test set. For each speaker (as a target speaker), we randomly select 50 utterances of this speaker and mix each of them with utterances from other speakers. As a result, we generate 95K, 3K, and 1.5K mixed utterances for training, validation and test, respectively. The signal-to-interference ratio (SIR) is equally distributed in [0dB, 6dB, 12dB, inf]. The number of speakers in the mixed signal ranges from 1 to 3 evenly. And the environmental noises are added to the speech mixture with signal-to-noise ratio (SNR) randomly sampled from [6dB, 12dB, 18dB, 24dB, 30dB].

\subsection{System setups}
For speaker embedding models, 257-d raw short time fourier transform (STFT) features are extracted with a 32ms window and the time shift of feature frames is 16ms. The non-speech part is removed by a critical-band power spectrum variance based voice activity detector (VAD), which provides better speech detection than the energy based VAD. The voiced part is randomly segmented into 100-200 frames for network training. A linear projection layer of 128-d is added to the TDNN and Resnet architecture for speaker embedding extraction, where $L_2$ norm is applied as an unit energy constraint. $\omega_1$=0.2 and $\omega_2$=0.001 in Eq.(\ref{eq3}) are determined in the experiments. 

We retain most of the hyper-parameters same with the target speaker enhancement model presented in \cite{ji2020}, except for the nontarget sample training. In our experiments, we find that the ratio 11:1 between target and nontarget samples are empirically good, which provides a balanced performance of target speaker extraction and interference speaker suppression. In the SV part of TASE-SVNet, we fine tune the SV model with enhanced speech. To prevent the fine-tuned ``spk embd net 2" from being over-fitted on the training data, we use a small learning rate of 1e-6 with just one epoch of fine-tuing.

\section{Results}

\label{results}
\subsection{Baseline SV results}

We first evaluate the baseline SV systems with the test corpora, i.e., Voxceleb, overlap-spk and vehicle-spk. We also reference the state-of-the-art Voxceleb benchmark results for fair comparison. 

\begin{table}[th]
\centering
\vspace{-0.2cm}
\caption{\label{sv} {\it SV baseline results (EER).}}
%\vspace{-0.2cm}
{\footnotesize{
\begin{tabular}{c|ccc}\toprule
  model & Voxceleb  & overlap-spk & vehicle-spk  \\
\cline{1-4}

 x-vector \cite{snyder2018x}& 3.14\%  & / & / \\ 	
 TDNN \cite{ji2020} & 2.10\%  & / &  / \\
 TDNN \cite{liu2019large} & 2.00\%  & / &  / \\
 TDNN (ours) & 1.95\%  & 16.32\% &  4.15\% \\
 Resnet-50 (ours) & 1.67\%  & 15.45\% &  3.72\% \\
 TDNN (ours T/S) & 1.76\%  & 15.91\% &  3.93\% \\
\bottomrule
\end{tabular}}}
\vspace{-0.2cm}
\end{table}
As illustrated on Table \ref{sv}, our proposed TDNN (6.5M) baseline could achieve comparable result as \cite{liu2019large} on Voxceleb. The large Resnet-50 model (31.4M) is able to approach the state-of-the-art single system performance reported in \cite{zeinali2019but}, where deeper Resnet models are used for speaker embedding training. For the ``overlap-spk" dataset, where only utterances with overlapped speakers are used for test, superb performance is no longer hold as the Voxceleb test set. For a more general purpose ``vehicle-spk" corpus, although performance degradation is not as big as the overlapping corpus, there still exists a big performance drop, showing the difficulties for SV in unconstrained environments. It is also noted that, the T/S trained TDNN model achieves consitently better performances across all tasks without any additional cost during inference. For the next experiments, the T/S trained TDNN model is therefore utilized in the TASE-SVNet framework.   

\begin{figure}[th]
  \centering

  \includegraphics[trim=3.2cm 0.8cm 7cm 0.6cm,width=60mm, height=60mm]{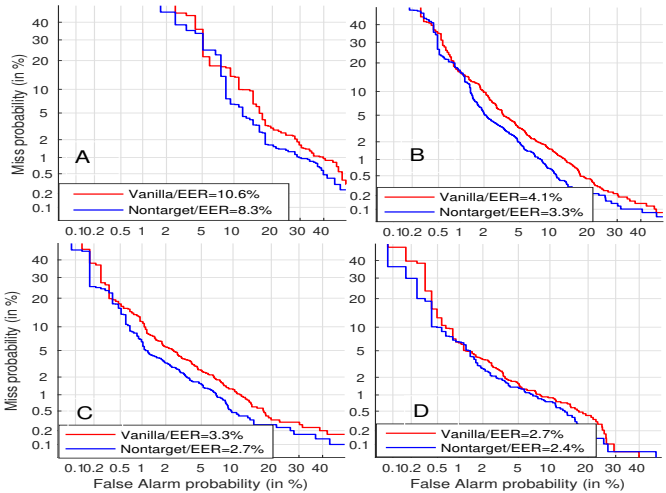}

  \caption{System performance comparison between the Vanilla TASE-SVNet and the nontarget sample TASE-SVNet on ``vehicle-spk“ test set, DET Curve is split based on 4 SNR conditions of the test utterances. The SNR condition in dB is: A=[-inf,3] B=[3,15], C=[15,20],D=[20, inf].   }
  \label{negative_fig}
  \vspace{-0.3cm}
\end{figure}

\subsection{Impact of nontarget sample training for TASE-SVNet}
In Sec.\ref{negative_samp}, we argue that conventional target speaker enhancement (i.e., the vanilla TASE-SVNet) models are trained with target samples and do not specifically address for nontarget speaker suppression. Therefore, a nontarget sample training constraint is added in TASE-SVNet. A comparison between the vanilla method and the nontarget sample version is given in Fig.\ref{negative_fig}. %From the spectrogram Fig.\ref{negative_fig} b), the vanilla target speaker enhancement module seems acting like a general noise suppression model, while the negative sample constrained target speaker enhancement Fig.\ref{negative_fig} c) performs much better in eliminating interference speakers. 
As illustrated in the DET curves, the nontarget sample based TASE-SVNet consistently outperforms the vanilla version. And the performance gain is from a lower FAR. We observe a lower FAR when the miss probability is fixed, which eventually achieves a lower EER. Table \ref{neg_eer} presents the overall SV performance across two test sets, as indicated in the table, the proposed nontarget sample based TASE-SVNet framework is robust in both highly overlapped condition and the general condition. The nontarget sample training strategy provides a +15.2\% relative improvement over the vanilla TASE-SVNet system in ``vehicle-spk" data. 
\begin{table}[th]
\centering
\vspace{-0.1cm}
\caption{\label{neg_eer} {\it SV results with nontarget sample TASE-SVnet.}}
%\vspace{-0.2cm}
{\footnotesize{
\begin{tabular}{c|cc}\toprule
  model &  overlap-spk & vehicle-spk  \\
\cline{1-3}
 TDNN (ours T/S)   & 15.91\% &  3.93\% \\
 TASE-SVNet (vanilla)   & 6.83\% &  3.72\% \\
 TASE-SVNet (nontarget)   & 6.02\% &  3.15\% \\
\bottomrule
\end{tabular}}}
\vspace{-0.3cm}
\end{table}

\subsection{Iterative inference}
One common issue remains for SV systems when the enrollment utterances are noisy, which impairs the system robustness in a long run. Since the TASE-SVNet framework is equipped with a target speaker enhancement module, it is feasible to speech enhancement for enrollment utterances as well. In fact, with the enhanced enrollment utterances, we can further perform target speaker extraction for test utterance iteratively.

The {\it{iterative inference}} that we conducted in our experiments stops after the second pass, for the consideration of performance and computational cost. The first pass is just directly inference with raw enrollment utterances. For the second pass, all 3 raw enrollment utterances are enhanced one by one, then the 3 enhanced enrollment utterances are employed as the anchor information for processing the test utterance. In our experiments, we find that combining the two passes will provide consistent performance gain.   

\subsection{Progressive system improvements}
We have presented the techniques that result in the improved SV performance. Table \ref{overall_eer} demonstrates the system improvements w.r.t different components. The best system achieves +63.8 and +30.3 relative improvements over a strong TDNN baseline, for overlap-spk and vehicle-spk test sets, respectively. Even compared with the TDNN fusion system, where similar computational consumption required for inference, the performance gain is considered significant. The improvement suggests that building the TASE-SVNet framework is essential generalization ability for SV systems in adverse acoustic environments.

\begin{table}[th]
\centering
\vspace{-0.2cm}
\caption{\label{overall_eer} {\it Progressive system improvements with the proposed techniques (EER).}}
%\vspace{-0.2cm}
{\footnotesize{
\begin{tabular}{l|cc}\toprule
  model &  overlap-spk & vehicle-spk  \\
\cline{1-3}
 TDNN (ours T/S)   & 15.91\% &  3.93\% \\
 TASE-SVNet (vanilla)   & 6.83\% &  3.72\% \\
 TASE-SVNet (+nontarget)   & 6.02\% &  3.15\% \\
 TASE-SVNet (+iterative)   & 5.89\% &  3.01\% \\
 TASE-SVNet (+two pass fusion)   & \textbf{5.75\%} &  \textbf{2.74\%} \\
 \toprule
 TDNN fusion (w/ + w/o T/S )   & 13.72\% &  3.65\% \\
\bottomrule
\end{tabular}}}
\vspace{-0.2cm}
\end{table}

\section{Conclusions}
\label{conclude}
In this study, we presented the TASE-SVNet for robust speaker verification. The main contributions of this study can be summarised as follows. We proposed to improve speaker embedding network with T/S training. State-of-the-art SV benchmark results were achieved with our proposed models. For speech enhancement front-end to SV models, we achieved balanced target speaker enhancement and nontarget speaker suppression. Our nontarget sample training strategy prevents the output of target speaker enhancement network from biasing to enrolled speaker for nontarget test utterance, which is essential in reducing FAR for SV. We implemented a flexible TASE-SVNet framework with multi-stage training. Iterative speaker enhancement is investigated to address noisy speaker enrollment, which relaxes the limitation of requiring clean speaker enrollment in conventional SV systems. Finally, significant performance gain on real SV corpora with noise and overlapped speech justified the effectiveness of our proposed framework.

\vfill\pagebreak

\bibliographystyle{IEEEbib}
\bibliography{refs}

\end{document}